%% file: main.tex
\documentclass[letterpaper,10 pt, conference]{ieeeconf}

\IEEEoverridecommandlockouts                              

\input{inc/packages}
\input{inc/definition_this_paper}
\makeatletter
\newcommand{\resetalgorithmlinenumber}{\setcounter{ALG@line}{0}}
\makeatother

\algnotext{EndFor}
\algnotext{EndIf}
\algnotext{EndWhile}

\usepackage{textcomp}

\makeatletter
\newcommand\fs@spaceruled{\def\@fs@cfont{\bfseries}\let\@fs@capt\floatc@ruled
  \def\@fs@pre{\vspace{10mm}\hrule height.8pt depth0pt \kern2pt}%
  \def\@fs@post{\kern2pt\hrule\relax\vspace{-10mm}}%
  \def\@fs@mid{\kern2pt\hrule\kern2pt}%
  \let\@fs@iftopcapt\iftrue}
\makeatother

\algrenewcommand\algorithmicfunction{\textbf{function}}
\algrenewcommand\algorithmicrepeat{\textbf{repeat}}
\algrenewcommand\algorithmicuntil{\textbf{until}}
\algrenewcommand\algorithmicreturn{\textbf{return}}
\algrenewcommand\algorithmicrequire{\textbf{Input:}}
\algrenewcommand\algorithmicprocedure{\textbf{procedure}}
\author{Md Abul Kashem Niloy$^1$, Adam Hallmark$^1$, Yikun Cheng$^2$, Pan Zhao$^1$,~\IEEEmembership{Member,~IEEE} 
\thanks{$^1$Md AK Niloy, A. Hallmark and P.~Zhao are with the Department of 
Aerospace Engineering and Mechanics, University of Alabama, Tuscaloosa, AL 35487, USA. Email:  \texttt{\small mniloy1@crimson.ua.edu, alhallmark@crimson.ua.edu,~pan.zhao@ua.edu}.}
\thanks{$^{ 2}$Y.~Cheng is with China University of Mining and Technology-Beijing, School of Mechanical and Electrical Engineering. Email: \texttt{\small yk.cheng@cumtb.edu.cn}.}
\thanks{\copyright~2026 IEEE. Personal use of this material is permitted. Permission from IEEE must be obtained for all other uses, in any current or future media, including  reprinting/republishing this material for advertising or promotional purposes, creating new collective works, for resale or redistribution to servers or lists, or reuse of any copyrighted component of this work in other works. Accepted for publication in \emph{IEEE Control Systems Letters}. DOI:~10.1109/LCSYS.2026.3713238}}

\def\BibTeX{{\rm B\kern-.05em{\sc i\kern-.025em b}\kern-.08em
    T\kern-.1667em\lower.7ex\hbox{E}\kern-.125emX}}
\begin{document}

\title{\LARGE \bf Neural-NPV Control: Learning Parameter-Dependent Controllers and Lyapunov Functions
}

\maketitle
\thispagestyle{empty}
\pagestyle{empty}

\begin{abstract}


This paper presents Neural-NPV Control, a learning-based framework for joint synthesis of a parameter-dependent (PD) controller and a PD Lyapunov function using neural networks for an NPV system under input constraints. At the first stage, the proposed framework utilizes a  gradient-based counterexample-guided procedure to synthesize a PD controller and a PD Lyapunov function candidate.  The second stage relies on a level-set guided procedure to refine the controller and Lyapunov function candidate while maximizing the robust region of attraction (R-ROA). The learned controller, Lyapunov function, and R-ROA are empirically evaluated. We demonstrate the advantages of Neural-NPV over SOS-based methods in terms of applicability,  performance, and scalability through numerical experiments involving  a simple inverted pendulum with one scheduling parameter and a quadrotor system with three scheduling parameters.
\end{abstract}



\section{Introduction}\label{sec:introduction}

A nonlinear parameter-varying (NPV) system is a nonlinear system whose behavior explicitly depends on external time-varying parameters. Many real-world nonlinear time-varying systems can be effectively represented as NPV systems. Examples include rockets with decreasing mass due to fuel consumption, ground vehicles subjected to variable friction coefficients, and aircraft experiencing disturbances that change over time. Similar to the linear parameter-varying (LPV) framework \cite{Rugh00gs_survey, Moh12LPVBook}, representing a nonlinear time-varying system as an NPV system facilitates the design of {\it gain-scheduled} nonlinear controllers that adjust the control strategy dynamically based on real-time measurements and/ or estimates of the scheduling parameters, while  avoiding the limitations of the LPV framework resulting from local validity or over-approximation of the LPV models \cite{Rugh00gs_survey}. 

Recent studies have used sum-of-squares (SOS) programming to synthesize nonlinear parameter-dependent (PD) controllers for NPV systems  \cite{fu2018hinf-npv,lu2020domain-npv,zhao2025parameter-pd-clf}. However, 
SOS-based methods suffer from limited scalability and structural restrictions for Lyapunov functions that potentially lead to conservative results. 
To overcome these limitations, this paper investigates the use of neural networks (NNs) to jointly learn PD controllers and PD Lyapunov functions for NPV systems.


\subsection{Related work}
\subsubsection{SOS optimization based nonlinear synthesis} SOS optimization \cite{parrilo2000structured-sos} has been widely adopted for stability analysis and controller synthesis of polynomial nonlinear systems. 
In the context of NPV systems, SOS-based methods have been applied to jointly synthesize nonlinear PD controllers and PD Lyapunov functions \cite{fu2018hinf-npv,lu2020domain-npv,zhao2025parameter-pd-clf}. Despite their theoretical appeal,  these methods suffer from two major  limitations: (i)  \textit{Restricted scalability}: 
 The computational complexity grows rapidly with the dimensionality of the system and the degree of the polynomial basis used, rendering them impractical for high-dimensional systems {\cite{ahmadi2019dSOS}}. 
(ii) \textit{Conservative performance due to structural restrictions} SOS-based nonlinear synthesis typically imposes structures on certificate functions to avoid a nonconvex problem, leading to potentially conservative performance. For example, SOS-based Lyapunov synthesis typically requires that the Lyapunov functions do not depend on states whose derivatives are directly affected by control inputs \cite{prajna2004nonlinear-sos,zhao2025parameter-pd-clf}.

\subsubsection{Learning certified control} There has been an increasing trend for learning a control certificate (alongside a controller) using  NNs\cite{dawson2023survey}, such as a Lyapunov function  \cite{dawson2023survey,chang2019neural-Lya}, or a contraction metric\cite{dawson2023survey,tsukamoto2020neural-contraction} among others. Unlike SOS-based methods that can only be applied to polynomial systems, certified neural control can be applied in the general nonlinear case  and is scalable to high-dimensional systems. 
 Recent advances further extend the learning-based paradigm to physics-informed formulations that characterize Lyapunov functions and enlarge the region of attraction via Zubov-type PDEs \cite{liu2025formally,li2026two}, and to tools integrating certificate learning with formal verification \cite{liu2024lyznet}. There has also been work to certify a controller for systems with unknown dynamics \cite{zhou2022neural}, where an NN is adopted to learn the unknown dynamics.
 
Nevertheless,  all of these works consider {\it time-invariant} dynamics. To the best of our knowledge, there is {\it no study on learning-based control for NPV systems.}

{\bf Our contribution}:
We propose Neural-NPV Control, a novel two-stage framework for jointly synthesizing  a PD Lyapunov function and a PD controller  using NNs for general NPV systems, unlike existing SOS-based methods that require 
control-affine, polynomial dynamics. Our framework can also handle asymmetric input constraints, which is challenging for SOS-based methods \cite{fu2018hinf-npv,zhao2023convex-cbf}. We evaluate the learned Lyapunov function and  controller through two empirical  evaluation schemes.  We show the superior performance of our framework in yielding a much larger robust region of attraction (R-ROA) compared to SOS-based methods using numerical experiments. We also demonstrate the scalability of our approach through numerical experiments on a quadrotor system with three scheduling parameters.

\begin{figure}[t]
    \centering    
    \includegraphics[width=1.0\columnwidth]{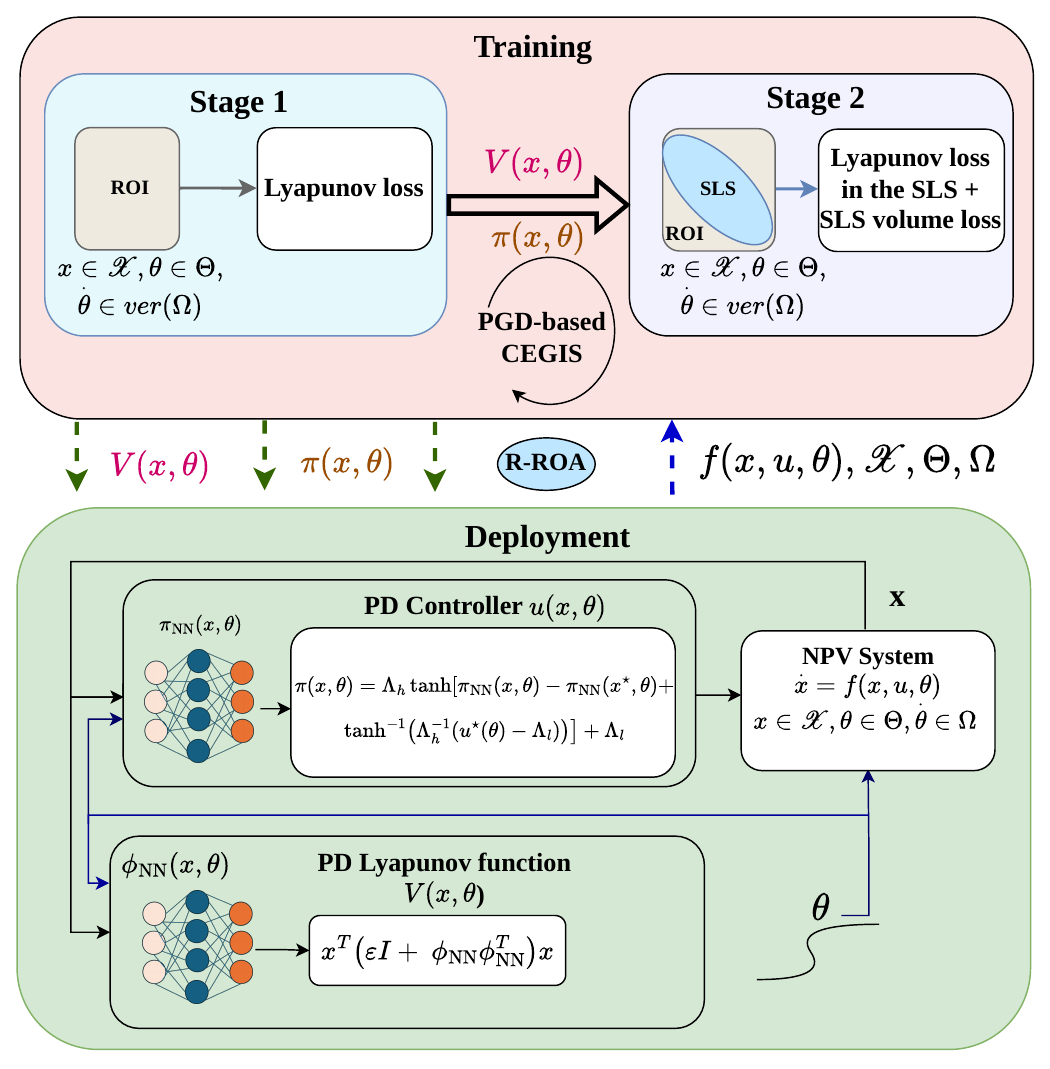}
    \vspace{-5mm}
    \caption{Neural-NPV framework. Within Neural-NPV, a PD controller $\pi(x,\theta)$ (to stabilize the NPV system) and a PD  Lyapunov function $V(x,\theta)$ (to certify the stability of the closed-loop system) are jointly trained to minimize the Lyapunov loss via supervised learning. ROI: Region of interest; ROA: Region of attraction
   }\label{fig:neural-npv framework}
\end{figure}



\section{Problem Settings and Preliminaries }\label{sec:preliminaries}
We consider a nonlinear parameter-varying (NPV) system,\begin{equation}
    \dot{x} = f(x, u, \theta),
    \label{eq:dynamics}
\end{equation}
where $x(t)\in \mbR^{n}$ is the state vector, $u(t)\in \mbR^{n_u}$  is input vector,  $\theta(t)\in\mbR^{n_\theta}$ denotes the time-varying parameter vector that can be measured online and $f:\mbR^n\times \mbR^\nu \times \mbR^\ntheta\ \rightarrow \mbR^n$  is locally Lipschitz in $x, u$ and $\theta$.  The parameter $\theta(t)$ characterizes the variation of the dynamics such as the mass of a rocket, which will be time-varying due to propellant consumption,
or external disturbance applied to an aircraft or a drone.  
The control input is subject to box constraints: 
\begin{equation}\label{eq:input-cst_}
\ubar{u} \leq u(t) \leq \bar u,\quad \forall t,
\end{equation}
where  $\ubar u\in\mbR^{\nu}$ and $\bar u\in \mbR^{\nu}$ denote the element-wise lower and upper bounds on $u$, respectively. 
\begin{assumption}\label{assump:theta-thetadot-bnd}
The time-varying parameters $\theta$ and their derivatives $\dot\theta$  satisfy\begin{equation} \label{eq:theta-thetadot-constrs}
\theta(t)\in \Theta \;\;\text{and}\;\; \dot{\theta}(t)\in {\Omega}, \quad \forall\, t\ge 0,
\end{equation}
where $\Theta$ is a compact set and $\Omega$ is a hyper-rectangular set defined by,\begin{equation}\label{eq:theta_assump_3}
\Omega \triangleq \bigl\{\, v \in \mathbb{R}^{n_\theta} \;:\; \underline{v} \le v \le \overline{v}\bigr\},  
\end{equation} 
where {$\ubar v \in\mbR^{n_\theta}$ and $\bar v\in \mbR^{n_\theta}$}. 
 For clarity, we define $\mathcal{F}_{\Theta}^{\Omega}$ as the set of all feasible 
parameter trajectories $\theta(t)$:\begin{equation}
  \mathcal{F}^{\Omega}_{\Theta}
  \triangleq
  \Bigl\{\theta:\mathbb{R}^{+}\to \mathbb{R}^{n_\theta}
  \;\big|\theta(t)\in\Theta,\dot{\theta}(t)\in \Omega,\; \forall\, t\ge 0 \Bigr\}.
\end{equation}
\end{assumption}

     Given \cref{eq:dynamics} subject to \cref{assump:theta-thetadot-bnd}, we are interested in designing a nonlinear PD controller      \begin{equation}\label{eq:control-law}
         u=\pi(x,\theta),
     \end{equation}
      that will stabilize the system \cref{eq:dynamics} under all admissible trajectories of $\theta$ and a PD Lyapunov function $V(x,\theta)$ certifying the stability. 
Under the controller \cref{eq:control-law}, the closed-loop system is given by  
\begin{equation}\label{eq:cl_dynamics}
\dot{x} = f(x,\pi(x,\theta), \theta)\trieq \bar f(x,\theta).
\end{equation}
Without loss of generality, we assume that  $x^\star = 0$ is an equilibrium state of the closed-loop system for all $\theta \in \Theta$.

\begin{definition} \cite[Definition 4.5]{khalil2002nonlinear-book}
The equilibrium point $x^\star = 0$ of the closed-loop system \cref{eq:cl_dynamics} is exponentially stable if there exist positive constants $c,~k$, and $\lambda$ such that 
 \begin{equation}{\label{eq:exp_equil}}
   \hspace{-3mm}    \norm{x(t)} \! \leq \! k \norm{x(t_0)}e^{-\lambda(t- t_0)}, \   \forall t\geq t_0 \! \geq 0, \ \forall \norm{x(t_0)}\!<\! c,
\end{equation}
for all admissible trajectories of $\theta$ satisfying \cref{eq:theta-thetadot-constrs}. 
%
\end{definition}

According to Lyapunov stability theory (as demonstrated in the proof of Theorem~1 in \cite{zhao2025parameter-pd-clf}), if there exists a PD controller $u = \pi (x,\theta)$ and a PD continuously differentiable  function  $V(x,\theta):  \mbR^{n_x} \times  \mbR^{n_\theta} \rightarrow \mbR $ and positive constants $k_1, k_2, k_3$ such that for all $x\in\mcX$ with $\mcX$ being a region of interest (ROI), the following conditions hold:
\vspace{-2mm}
\begin{subequations} \label{eq:exp_condition}
    \begin{align} 
    V(0,\theta) = 0, \quad \forall \theta \in \Theta, \label{eq:exp_condition_a} \\ 
    k_{1}\|x\|^{2} \leq V(x,\theta) \leq k_{2}\|x\|^{2}, \quad  \forall \theta \in \Theta, \label{eq:exp_condition_b} \\    
        \dot{V}(x,{\theta}) 
     \leq -k_{3}\|x\|^{2},         \quad \forall \theta \in \Theta, \; \forall \dot{\theta} \in \textup{Ver}(\Omega), \label{eq:exp_condition_c}
    \end{align}
\end{subequations}
  where 
  \begin{equation}\label{eq:V-derivative}
      \dot{V}(x,{\theta}) 
       \! = \!\frac{\partial V}{\partial x}\cdot \bar f +  \frac{\partial V}{\partial \theta}\cdot \dot\theta 
       , 
  \end{equation}
 with $\bar f$ defined in \cref{eq:cl_dynamics}, and
 $\textup{Ver}(\Omega)$ denotes the set of vertices of $\Omega$, then the closed-loop system \cref{eq:cl_dynamics} is exponentially stable. 

\begin{remark}
Since \cref{eq:exp_condition_c} is affine with respect to $\dot \theta$, if it holds for all  $\dot \theta \in \textup{Ver}(\Omega)$, then it holds for any $\dot \theta \in \Omega$.
\end{remark}

\begin{definition}
The robust region of attraction (R-ROA) for the closed-loop system
\cref{eq:cl_dynamics} is the set of all initial states for which the closed-loop trajectory
satisfies $\lim_{t\to \infty} x(t)=0$ for all admissible
trajectories of $\theta$. Formally, an R-ROA is a set defined by
\begin{equation}
\mathcal{A}= \left\{
x(0)\in\mathbb{R}^n:\ \forall \theta\in\mathcal{F}^{\Omega}_{\Theta},\ 
\lim_{t\to \infty} x(t)=0
\right\}.  \label{eq:r-roa}
\end{equation}
\end{definition}

\section{Joint Learning of PD Controllers and Lyapunov functions}

%
We now present a method to jointly learn a PD controller and PD Lyapunov function 
using NNs that (approximately) satisfy the conditions \cref{eq:exp_condition}.

\subsection{Parameterizing the PD controller and Lyapunov function}\label{sec:lyapunov-parameterization}
The Lyapunov function should meet all the conditions in \cref{eq:exp_condition}. To simplify the training process, we parameterize our Lyapunov function in a way that conditions \cref{eq:exp_condition_a} \& \cref{eq:exp_condition_b} can be automatically satisfied. Inspired by \cite{yang2024lyapunov}, we adopt the following parameterization:\begin{equation}\label{eq:lyapunov} V(x,\theta) = x^T\left( \phi_\textup{NN}(x,\theta)\phi_\textup{NN}^T(x,\theta)+\epsilon I \right)x,
\end{equation}where $\epsilon$ is a small positive constant, $I$ is the identity matrix, and $\phi_\textup{NN}(x,\theta)$ is an NN with $n+n_\theta$ neurons in the input layer and $n$ neurons in the output layer. $\phi_\textup{NN}(x,\theta)$ takes both state $x$ and parameters $\theta$ as the input.  We use \texttt{tanh} activation function for $\phi_\textup{NN}$, which allows the computation of the derivative of $V(x,\theta)$ defined in \cref{eq:V-derivative}. 

With the parameterization in \cref{eq:lyapunov}, \cref{eq:exp_condition_a} is automatically satisfied. To see that \cref{eq:exp_condition_b} is also satisfied by construction, first note that $ \phi_\textup{NN}(x,\theta)\phi_\textup{NN}^T(x,\theta)$ is positive semi-definite, since it is the outer product of a real vector with itself. Let $\bar \lambda\geq 0$ and $\ubar \lambda\geq 0$ denote the largest and smallest eigenvalues, respectively, of $ \phi_\textup{NN}(x,\theta)\phi_\textup{NN}^T(x,\theta)$ over all $x\in\mcX$ and $\theta\in\Theta$. Then, we have  
\begin{equation}\label{eq:V-bnd-lambda}
    (\ubar\lambda +\epsilon)\norm{x}^2\leq V(x,\theta) \leq (\bar\lambda +\epsilon)\norm{x}^2,\ \forall x\in \mcX, \theta\in\Theta.
\end{equation}
Setting $k_1= \ubar\lambda +\epsilon>0$ and $k_2= \bar\lambda +\epsilon>0$, we see that the parameterization \cref{eq:lyapunov} satisfies \cref{eq:exp_condition_b} by construction. 
To represent the controller, we use a multi-layer NN $\pi_{\textup{NN}}(x,\theta)$ with $n+n_\theta$ neurons in the input layer and  $\nu$ neurons in the output layer. 
We denote  the vectors of trainable  parameters of $\phi_\textup{NN}(x,\theta)$ and $\pi_{\nn}(x,\theta)$ as $W_V$ and $W_\pi$, respectively.
 To guarantee that $x^{\star} = 0$ remains a closed-loop equilibrium for every $\theta \in \Theta$ and that the input respects the physical bounds $\underline{u} \leq u \leq \bar{u}$ we parameterize the controller as
\begin{multline}\label{controlpolicy}
\pi(x,\theta) = \Lambda_h \tanh \Bigl[\, \pi_{\textup{NN}}(x,\theta) - \pi_{\textup{NN}}(x^{\star},\theta) \\
{}+ \tanh^{-1}\!\Bigl(\Lambda_h^{-1}\bigl(u^{\star}(\theta) - \Lambda_l\bigr)\Bigr) \,\Bigr] + \Lambda_l,
\end{multline}
where $\Lambda_h = \mathrm{diag}\bigl((\bar{u} - \underline{u})/2\bigr)$ 
is the diagonal half-range matrix, 
$\Lambda_l = \tfrac{1}{2}(\bar{u} + \underline{u})$ 
is the midpoint vector, $\tanh$ and $\tanh^{-1}$ act component-wise, and $u^{\star}(\theta)$ is 
the trim input satisfying 
$\bar{f}(x^{\star}, u^{\star}(\theta), \theta) = 0$, for any $\theta \in \Theta$. 
Two properties of~\eqref{controlpolicy} hold by construction, i.e., for 
any setting of the network weights: (i)~the residual 
$\pi_{NN}(x, \theta) - \pi_{NN}(x^{\star}, \theta) = 0$ at 
$x = x^{\star}$, so 
$u(\theta) = \Lambda_h \Lambda_h^{-1}\bigl(u^{\star}(\theta) - \Lambda_l\bigr) + \Lambda_l = u^{\star}(\theta)$ 
for every $\theta \in \Theta$, and hence $x^{\star}$ is a closed-loop 
equilibrium uniformly over $\Theta$; (ii)~since $\tanh \in (-1, 1)^{n_u}$ 
component-wise and $\Lambda_h$ is positive diagonal, 
$\pi(x, \theta) \in \Lambda_l + \Lambda_h \cdot (-1, 1)^{n_u} = 
(\underline{u}, \bar{u})$ for every $(x, \theta)$. 
The form generalizes the saturated residual parameterization of 
\cite{li2026two} to NPV systems with $\theta$-dependent trim.
The effectiveness of this simple strategy for enforcing input constraint is validated in \cref{sec:sim}.



\subsection{The Neural-NPV Control algorithm}
 Within Neural-NPV, we jointly train the Lyapunov network $\phi_{\nn}$ and controller network $\pi_{\nn}$ using a two stage procedure. In {\bf Stage I}, we use counterexample guided synthesis to obtain a working controller and a Lyapunov function candidate over the full domain $\mathcal{X}\times\Theta$, where $\mcX$ is the region of interest. In {\bf Stage II}, we refine the controller and Lyapunov networks using a level-set-guided loss to  estimate the R-ROA and a surrogate loss that promotes the growth of the R-ROA. 
 The algorithm is summarized in \cref{neuralNPV}. 

\subsubsection{Stage I: Counterexample-guided joint synthesis}

Existing work on learning certified control typically first generates samples from the state space, uses these samples to evaluate the violation of the  Lyapunov condition \cref{eq:exp_condition}, and optimizes the network parameters to minimize the violation. 
We follow this idea. However, here, we  sample from the space of ($x$, $\theta$) i.e., $\mcX \times \Theta$, and enumerate $\dot \theta$ in  Ver$(\Omega$). At each iteration, we generate $N$ number of  $x$-$\theta$ pairs, and then use CEGIS (Counterexample guided Inductive Synthesis) framework \cite{yang2024lyapunov}  to iteratively generate adversarial samples $(x^i_\textup{ad},\theta^i_\textup{ad})$.  Instead of using expensive verifiers such as mixed-integer programming (MIP) \cite{dai2021lyapunov}, or satisfiability modulo theories (SMT) \cite{chang2019neural-Lya} to generate counterexamples, we use projected gradient descent (PGD) \cite{yang2024lyapunov,li2026two} to generate the worst adversarial examples that maximize the following objective:  \begin{equation}\label{eq:pgdcost}
\hspace{-4mm}L_{\dot V}(x^i_\textup{ad}, \theta^i_\textup{ad}) = \max_{\dot\theta \in \textup{Ver}(\Omega)}\! \dot{V}(x^i_\textup{ad}, \theta^i_\textup{ad}) + k_3 \|x^i_\textup{ad}\|^2.
\end{equation}
By maximizing \cref{eq:pgdcost}, we aim to find adversarial samples that have the maximum violation of the condition \cref{eq:exp_condition_c}. Since \cref{eq:exp_condition_a,eq:exp_condition_b} have already been satisfied by construction as explained in \cref{sec:lyapunov-parameterization}, we only need to evaluate \cref{eq:exp_condition_c}. At each $(x^i_\textup{ad},\theta^i_\textup{ad})$, we evaluate $V(x^i_\textup{ad},\theta^i_\textup{ad})$ and $u(x^i_\textup{ad},\theta^i_\textup{ad})$ and thus assess whether the  conditions in \cref{eq:exp_condition} are violated at  $(x^i_\textup{ad},\theta^i_\textup{ad})$ and if so, the level of violation. These adversarial examples are added to a buffer and then used to optimize the following empirical Lyapunov loss:
\vspace{-3mm}
 \begin{equation}
\label{eq:coststage1}
\hspace{-5mm}L_{\mathrm{lya}}\!=\! \frac{1}{N N_v}\! \sum_{i=1}^{N} \sum_{\dot\theta \in \mathrm{Ver}(\Omega)} \!\!\!\!\mathrm{ReLU}\!\left(\, \dot{V}(x^i_{\textup{ad}}, \theta^i_{\textup{ad}})\big|_{\dot{\theta}} +\! k_3\,\|x^i_{\textup{ad}}\|^2 \,\right),
\vspace{-5mm}
\end{equation}
with $N_v = |\mathrm{Ver}(\Omega)|$,
which characterizes the violation of \cref{eq:exp_condition_c} using the $N$ samples. 
We minimize the Lyapunov loss in \cref{eq:coststage1} to jointly learn a working controller and a  Lyapunov function candidate over the ROI $\mathcal{X} \times \Theta$. 
We stop once we see  the
counterexample violation rate drops below $1\%$ and the counterexample buffer size plateaus for 100 consecutive iterations, indicating
that the network satisfies condition \cref{eq:exp_condition_c} over the vast
majority of the sampled state-disturbance space.
\begin{algorithm}[t]
\caption{ Neural-NPV Control}\label{alg:NeuralLyapunovjan26}
\begin{algorithmic}[1]
\Require Lyapunov network $\phi_{\mathrm{NN}}(\cdot)$; Control network $\pi_{\mathrm{NN}}(\cdot)$; Learning rate $\eta$; ROI $\mathcal{X}$; Parameter sets $\Theta$ and $\Omega$; Counterexample (CEX) buffer $\mathcal{D}$; Candidate sample buffer $\mathcal{C}$; Maximum iteration $N_{\textup{iter}}$; Maximum epoch $N_{\textup{epoch}}$; PGD step $N_{\textup{PGD}}$; Max stagnation $n_s$; Samples per iteration $S$; Constant $k_3$; PGD step size $\beta$; level-set constant $\rho$ %
\vspace{5pt}

\Statex \textbf{Stage I: Counterexample-guided joint synthesis}
\State Initialize $\mathcal{D}$ $\leftarrow \emptyset$
\For{iter = 1:$N_{\textup{iter}}$}
\State Sample randomly $S$ points ($x_{\textup{ad}},\theta_{\textup{ad}}) \in (\mathcal{X} \times \Theta$)
    \For{$i = 1:N_{\textup{PGD}}$}
        \State Maximize violation via gradient ascent:
        \State  ($x_\textup{ad},\theta_\textup{ad}) \leftarrow \text{Project}_{(\mcX\times\Theta)}\Bigl((x_\textup{ad},\theta_\textup{ad})$
        
        $+ \beta \nabla_{x_\textup{ad},\theta_\textup{ad}}L_{\dot V}(x_\textup{ad},\theta_\textup{ad})\Bigr)$ $\triangleright$ See eq. \cref{eq:pgdcost} 
    \EndFor
\State Update counterexample buffer, $\mathcal{D} \leftarrow \mathcal{D} \cup (x_{\textup{ad}},\theta_{\textup{ad}})$
    \For{j = 1:$N_{\textup{epoch}}$}  
        \State Compute $L_\textup{lya}$  in \cref{eq:coststage1} using all samples in $\mathcal{D}$
        \State $W_V,W_\pi \;\leftarrow\; W_V,W_\pi \;-\;\eta\,\nabla_{(W_V,W_\pi)}{L}_\textup{lya}$
    \EndFor
\EndFor
\Statex\textbf{Output} $\phi_{\nn}$ \& $\pi_{\nn}$
\\

\Statex \textbf{Stage II: Level-set-guided ROA refinement}
\resetalgorithmlinenumber
\State Initialize $\mathcal{D}$ $\leftarrow \emptyset$
\For{iter = 1:$N_{\textup{iter}}$}
    \State Sample randomly $S$ points ($x_{\textup{ad}},\theta_{\textup{ad}}) \in (\mathcal{X} \times \Theta$)
    \For{i = 1:$N_{\textup{PGD}}$}
        \State Maximize violation via gradient ascent in the level-set:
        \State  ($x_\textup{ad},\theta_\textup{ad}) \leftarrow \text{Project}_{(\mcX\times\Theta)}\Bigl((x_\textup{ad},\theta_\textup{ad})$
        
        $+ \beta \nabla_{x_\textup{ad},\theta_\textup{ad}}L_{\dot V; \rho}(x_\textup{ad},\theta_\textup{ad}; \rho)\Bigr)$   $\triangleright$ See eq. \cref{loss-levelset}
    \EndFor
    \State Update counterexample buffer, $\mathcal{D} \leftarrow \mathcal{D} \cup (x_{\textup{ad}},\theta_{\textup{ad}})$
    \State Sample randomly $S$ points $x_{\cand} \in \mathcal{X}$, and append to buffer $\mathcal{C}$ $\leftarrow$ $x_{\cand}$
    \For{j = 1:$N_{\textup{\textup{epoch}}}$}  
        \State Update model parameters by optimizing eq. \cref{cost:stage2}
        \State $ W_V,W_\pi \leftarrow {W_V,W_\pi} - \eta \nabla_{(W_V,W_\pi)} L_{total}$
    \EndFor
    \If{CEX violation rate $<$ 1\% and R-ROA growth is stagnated for $n_s$ iterations}
        \State BREAK
    \EndIf
\EndFor
\Statex\textbf{Output} $\phi_{\nn}$ \& $\pi_{\nn}$
\label{neuralNPV}
\end{algorithmic}
\end{algorithm} 

\subsubsection{Stage II: Level set-guided ROA refinement}

Stage I aims to get a working controller and a Lyapunov function for the whole ROI. In Stage II, we introduce  a level set 
\vspace{-2mm}
\begin{equation}
     \Lambda^{\rho}_V = \left\{ x \in \mathbb{R}^n :\max_{\theta \in \Theta}\, V(x, \theta) \leq \rho \right\}, \label{eq:levelset-roa}
\end{equation}
in the training process, and refine the Lyapunov function candidate and the controller together, so that the level set represents the R-ROA, and furthermore its size is maximized. In practice, we can fix the constant $\rho$, e.g., to 1, since we can always update $V(x,\theta)$ to adjust the level set. 

   For $\Lambda^{\rho}_V$ to be a valid R-ROA, the Lyapunov conditions defined in \cref{eq:exp_condition} must be satisfied for all states inside $\Lambda^{\rho}_V$ and for all $\theta \in \Theta$. Given our parameterization, we only need to focus on \cref{eq:exp_condition_c}. Thus, we define the following objective:
   \vspace{-2mm}
 \begin{equation} \label{loss-levelset}
   \hspace{-2mm} L_{\dot V ; \rho }\!= \min \left ( L_{\dot V}(x^i_\textup{ad},\theta^i_\textup{ad}), \rho \! -\! \max_{\theta \in \Theta}V(x^i_\textup{ad}, \theta)\right).
\end{equation}
We still follow the counterexample-guided training process and find adversarial examples in the level-set $\Lambda^{\rho}_V$ with a PGD attack which violate \cref{loss-levelset}.  In implementation, $\max_{\theta \in \Theta} V(x, \theta)$ is approximated by evaluating $V(x,\cdot)$ on a uniform grid of $\theta$ over $\Theta$ and taking the maximum, yielding 27 evaluations per state for the quadrotor and 10 for the inverted pendulum. This resolution is sufficient because $V(x, \cdot)$ is Lipschitz in $\theta$ over the compact set $\Theta$, owing to tanh activations and finite weights of $\phi_{\mathrm{NN}}$. By concentrating on the level-set, we are able to significantly reduce the number of counterexamples and are able to find the R-ROA for a given NPV system.  Since the level-set value $\rho$ is fixed during the training procedure, the R-ROA found can be very conservative. In order to find the R-ROA that covers a large portion of the state space, we introduce a surrogate loss
 \vspace{-2mm}
\begin{equation}L_{\Lambda^{\rho}_V} = \frac{1}{N} \sum_{i=1}^{N}  \max_{\theta \in \Theta} V(x^i_\textup{\cand}, \theta) - \rho. 
    \label{loss-roa} 
\end{equation}
Equation \cref{loss-roa} indirectly promotes the growth of the level-set $\Lambda^{\rho}_V$. We sample uniformly from the state space to get $x_{\cand}$ that are used to compute \cref{loss-roa}. This surrogate loss ensures the robust level-set $\Lambda^{\rho}_V$ does not shrink to a neighborhood of the origin, and  covers the region of interest as much as possible.  The joint cost function of this refinement stage with objective weights $\beta_1$ and $\beta_2$ is\begin{equation}
\vspace{7mm}\label{cost:stage2}
L_{total} = \beta_1 L_{\mathrm{lya}} + \beta_2 L_{\Lambda^{\rho}_V},
\end{equation} where the first objective tries to certify the level-set as a valid subset of the R-ROA using adversarial samples in the level-set while the second one tries to maximize its volume.

 \subsubsection{Initialization of the control network}
Before entering Stage I, we initialize our PD controller to enhance the training process. This is to avoid convergence issues when starting with a randomly initialized controller network. Inspired by \cite{chang2019neural-Lya}, we first design a gain-scheduled linear state-feedback law based on LQR using local linearization at various $\theta$. We then train the initial controller network to match the gain-scheduled controller, i.e., we minimize the loss $||\pi_{\nn}(x,\theta) - K(\theta)x||^2$ over $(x,\theta)$ samples so that the PD controller has a reasonable starting point.

\subsection{Empirical  evaluation}
\label{sec:verification}
 In order to formally verify the Lyapunov conditions for our NPV systems over the estimated R-ROA, we need to check \cref{eq:exp_condition_c} for all admissible $(x,\theta,\dot\theta)$, where $(x,\theta) \in (\mathcal{X}\times\Theta)$, $\dot\theta \in \mathrm{Ver}(\Omega)$, and $x$ lies inside the robust level set $\Lambda_V^\rho$ defined in~\cref{eq:levelset-roa}. Existing neural Lyapunov verifiers, such as $\alpha,\beta$-CROWN \cite{li2026two, yang2024lyapunov} and LyZNet \cite{liu2024lyznet}, can in principle verify properties over the joint input box $(x,\theta,\dot\theta)$. However, applying them to \cref{eq:exp_condition_c} over the restricted set $\Lambda_V^\rho$ is not straightforward, because the verifier would also need to enforce the condition $(\max_{\theta \in \Theta} V(x,\theta) \le \rho)$ as a precondition. This type of nested robust level-set constraint is not directly supported by these tools in their current form. We therefore adopt two empirical evaluation schemes, described below.
\subsubsection{PGD-based  evaluation}
We run a strong adversarial PGD attack on our trained model, aiming to find states in the level-set $\Lambda^{\rho}_V$ that  violate the Lyapunov derivative condition \cref{eq:exp_condition_c} (with a tolerance) for some $\theta \in \Theta$ and at any $\dot\theta \in  \mathrm{Ver}(\Omega)$. We generate $N_\text{adv}$ adversarial samples from this attack and consider our controller and Lyapunov function  to pass this evaluation when the violation rate, i.e., the ratio of samples that violate the condition~\cref{eq:exp_condition_c} over all samples, remains below a small number, e.g., $1\%$, consistently.
\subsubsection{Trajectory-based evaluation}
We sample $10^6$ points from the level-set $\Lambda^{\rho}_V$ and  roll out closed-loop trajectories under time-varying $\theta(t)$. For each trajectory, we  check that the state converges to $x^\star = 0$ and that the Lyapunov function decreases monotonically,  providing empirical evidence that $\Lambda^\rho_V$ is a positively invariant set under the learned control law.

\section{Simulation Results}\label{sec:sim}

We demonstrate the performance and scalability of our proposed methods through numerical experiments using two systems: a simple inverted pendulum and a quadrotor.  The codes are available at \url{https://github.com/cal-research/Neural-NPV-Control}.

\subsection{Inverted pendulum with varying control effectiveness}
The dynamics of an inverted pendulum are given by
\begin{equation}
   \dot x(t) = \begin{bmatrix}
       \dot \phi (t)\\
      \frac{g}{L} \sin(\phi (t)) - \frac{b}{mL^2}\dot \phi (t)
   \end{bmatrix} 
   + \theta(t)  \begin{bmatrix}
       0 \\
       \frac{1}{mL^2}
   \end{bmatrix} u(t),
\end{equation}
where $x=[\phi, \dot \phi]^T$  and $u = \tau$ denote the state and control input (torque), respectively,   $m=0.1$, $L=0.5$, $b=0.2$, and $g=9.81$ denote the mass, length, damping ratio, and gravitational constant respectively. We impose an input constraint of $\abs{u(t)}\leq 3$ for all $t\geq 0$.
The parameter $\theta(t)$ is introduced to characterize the actuator effectiveness. 

 We consider the following ROI: $[\phi,\dot\phi] \in \mathcal{X}=[-\pi,\pi] \times [-6,6]$, $\theta(t)\in\Theta = [0.2,1]$ and $|\dot\theta(t)|\leq 0.1$. The structure of the Lyapunov and the controller network is shown in  \cref{tab:hyperparams}. Notice that both  networks have four input channels rather than three. In order to handle  the discontinuity at $\phi = \pm\pi$ and provide a
smooth, globally valid state representation for the network, we convert the angular state $\phi$ to $(\sin\phi,\, \cos\phi)$, which is fed to both control and Lyapunov networks together with $\dot\phi$ and $\theta$. 
\begin{table}[t]\
    \centering
    \caption{Network architectures and  hyperparameters}
    \label{tab:hyperparams}
    \renewcommand{\arraystretch}{1.2}
    \begin{tabular}{|l|c|c|c|c|}
        \hline
        System & $\phi_{\mathrm{NN}}$ dimension & $\pi_{\mathrm{NN}}$ dimension & $\rho$ & $\eta$ \\
        \hline
        Inv. Pendulum & [4, 64, 128, 3] & [4, 64, 128, 1] & 1 & $10^{-5}$ \\
        \hline
        3D Quadrotor & [9, 64, 128, 6] & [9, 64, 128, 64, 3] & 1 & $10^{-5}$ \\
        \hline
    \end{tabular}
\end{table}

\begin{figure}[ht]
\vspace{-3mm}
\begin{minipage}[t]{0.6\columnwidth}
 \centering
\includegraphics[width=\linewidth]{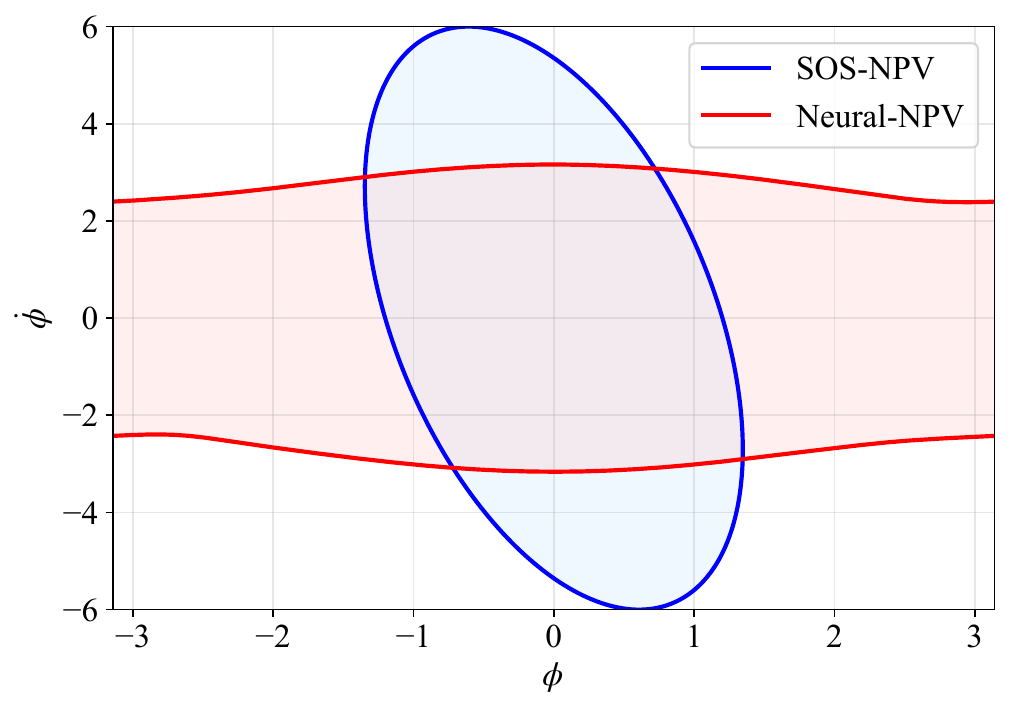}
\end{minipage}
\begin{minipage}[t]{0.38\columnwidth}
    \vspace{-32mm}
    \caption{Estimated R-ROA under SOS-NPV and Neural-NPV for the inverted pendulum system}
    \label{fig:roa-pendulum}
    \end{minipage}
    \vspace{-4mm}
\end{figure}

We compared Neural-NPV to SOS-NPV \cite{zhao2025parameter-pd-clf},
which jointly synthesizes a PD controller and PD Lyapunov function
via SOS optimization, extended here with input constraints
following \cite{zhao2023convex-cbf}. From \cref{fig:roa-pendulum}, we can see that Neural-NPV achieves the R-ROA whose volume is
markedly larger than that obtained by SOS-NPV, and covers the whole state space along the $\phi$ axis.
In SOS-NPV, the Lyapunov function was parameterized as $V(x,\theta) = x^TX^{-1}(\phi,\theta)x$, where $X(\phi,\theta)$ is a  5th-order polynomial matrix function. Notice that we intentionally made $X(\phi,\theta)$  not dependent on $\dot\phi$ (whose derivative is directly affected by $u$) to have a convex optimization problem \cite{prajna2004nonlinear-sos,zhao2025parameter-pd-clf}. In comparison, within Neural-NPV, the matrix term in the Lyapunov function \cref{eq:lyapunov}   has a more general structure (not limited to the inverse of a polynomial function) and can depend on all states. This demonstrates the superiority of our framework compared to SOS-based optimization methods.

To validate the estimated R-ROA, we first tried the PGD-based evaluation scheme detailed in \cref{sec:verification}. For this, we generated $10^6$ adversarial samples in $\Lambda^{\rho}_V \times \Theta$ and obtained a violation rate of approximately $1\%$ with a tolerance of $10^{-3}$. We next applied the trajectory simulation based evaluation scheme also detailed in \cref{sec:verification}. For this, we simulated the system in closed loop under a time-varying disturbance trajectory defined by $\theta (t) = 0.6 + 0.4 \cos (0.25t)$, and $10^6$ sampled initial states from the R-ROA and ROI. We found that the closed-loop transitions converge to the equilibrium across all the simulated cases in both tests.  The results are summarized in \cref{tab:verification_results}, which not only  empirically validate $\Lambda^{\rho}_V$ as a subset of the R-ROA, but also validate the performance of the controller in the entire ROI. 

\begin{table}[ht]\
    \centering
    \caption{ Empirical evaluation of estimated R-ROA and the learned controller}
    \label{tab:verification_results}
    \footnotesize
    \renewcommand{\arraystretch}{1.3}
    \setlength{\tabcolsep}{4pt}
    \begin{tabular}{|l|l|c|c|c|}
        \hline
        System & Scheme & Samples & Violation rate & Convergence  \\
        \hline
        \multirow{2}{*}{Inv.\ Pendulum}
            & PGD        & $10^6$ & $~1\%$  & --    \\
            & Trajectory & $10^6$ in  $\Lambda_V^\rho$ & --     & 100\%   \\
              & Trajectory & $10^6$ in ROI & --     & 100\%   \\
        \hline
        \multirow{2}{*}{Quadrotor}
            & PGD        & $10^6$ & $1\%$  & --    \\
            & Trajectory & $10^6$ in  $\Lambda_V^\rho$ & --     & 100\%   \\
            & Trajectory & $10^6$ in ROI & --     & 100\%   \\
        \hline
    \end{tabular}
    \vspace{-2mm}
\end{table}

\subsection{Quadrotor with varying external disturbance} \label{subsec:quad-dynam}
 To demonstrate the scalability of Neural-NPV to high-dimensional systems, we apply it to a quadrotor system subject to external disturbances: 
 \begin{equation}
    \hspace{-3mm}\begin{bmatrix} \ddot{p}_x \\ \ddot{p}_y \\ \ddot{p}_z \end{bmatrix}
    \!\!=\! g{e}_3 - \frac{\tau}{m} R e_3 + \frac{\theta}{m}
    \!=\! \!\begin{bmatrix} -\frac{\tau}{m}\cos\phi\sin\theta_p  \!+\! \frac{\theta_x}{m} \\ \frac{\tau}{m} \sin\phi \!+\! \frac{\theta_y}{m} \\ g - \frac{\tau}{m}\cos\theta_p\cos\phi \!+\! \frac{\theta_z}{m} \end{bmatrix},
    \label{eq:quad_translational}
    \vspace{-8mm}
\end{equation}
where the state vector is $x = \begin{bmatrix} p_x,\ p_y,\ p_z,\ \dot{p}_x,\ \dot{p}_y,\ \dot{p}_z\end{bmatrix}^\top \in \mathbb{R}^{6}$, $g=9.81 m/s^{2}$, ${e}_3 = [0,0,1]^\top$, and $R$ is the rotation matrix from body frame to the inertial frame following a ZYX (yaw-pitch-roll) sequence. The parameter $\theta (t) = \begin{bmatrix}\theta_x,\theta_y, \theta_z
\end{bmatrix} \in \mbR^3$ denotes the external disturbance forces along $x$, $y$, and $z$ axes. Here, we assume that the disturbance is known. In practice, the disturbances can be estimated, e.g., using an adaptive estimator \cite{wu2025L1quad}. 
The attitude (roll, pitch, yaw) is parameterized as $(\phi, \theta_p, \psi)$, and $\tau > 0$ denotes the total thrust  generated by the four rotors. The control input is defined as $
    {u} \triangleq \begin{bmatrix}  \tau,\ \phi,\ \theta_p \end{bmatrix}^\top$. The outer-loop controller outputs desired thrust $\tau$ and attitude angles $(\phi, \theta_p)$, which serve as references
for an inner-loop attitude controller assumed to operate on a much faster timescale. As yaw is not actively controlled, we set $\psi = 0$.  Since  the dynamics in \cref{eq:quad_translational} are not affine in control, we cannot directly apply SOS-based methods, which are only applicable to control-affine systems \cite{fu2018hinf-npv,lu2020domain-npv,zhao2025parameter-pd-clf}.

\begin{figure}[h]
\begin{minipage}[t]{0.55\columnwidth}
 \centering
\includegraphics[width=1\columnwidth]{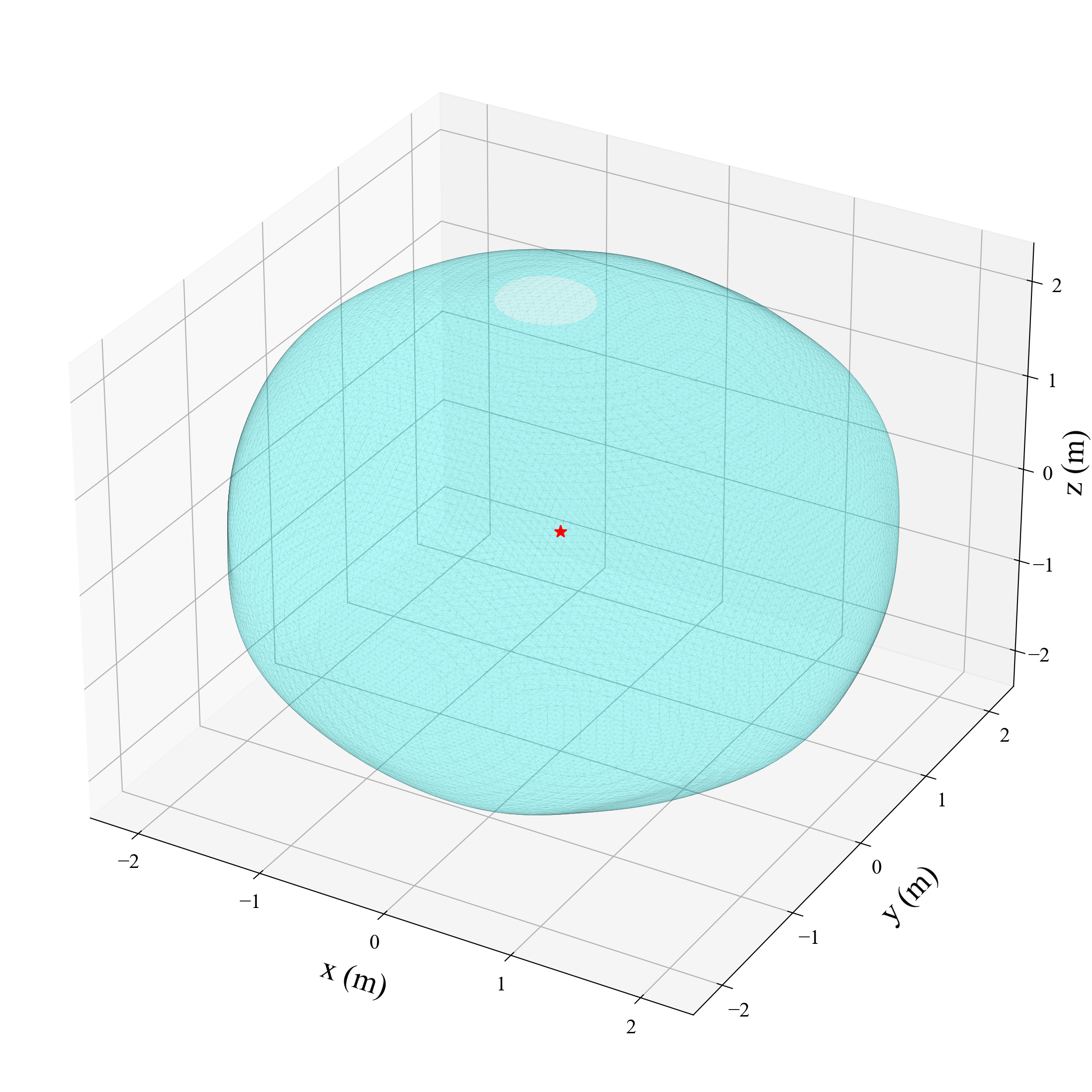} 
\end{minipage}
\begin{minipage}[t]{0.35\columnwidth}
    \vspace{-30mm}
   \caption{3D slice of the estimated R-ROA $\Lambda^{\rho}_V$ for the quadrotor in $x$-$y$-$z$ space.}
    \label{fig:quad-roat}
    \end{minipage}
\end{figure}

The following input constraints are imposed: $(\phi,\,\theta_p) \in [-90^\circ,\,90^\circ]\times[-180^\circ,\,180^\circ], \tau \in [0\,,\,11.3]$ N. We consider the following ROIs: $x \in \mathcal{X} = [-6,6]^6, \theta (t) \in \Theta = [-2,2]^3,\ \dot \theta (t) \in \Omega = [-0.5, 0.5]^3$. \cref{tab:hyperparams} lists the details of the Lyapunov network and the controller network, which are trained following \cref{alg:NeuralLyapunovjan26}. The estimated R-ROA 
is illustrated in \cref{fig:quad-roat}.  To empirically assess the candidate R-ROA, we first applied the PGD-based evaluation described in \cref{sec:verification}. For this, we generated $10^6$ adversarial samples in $\Lambda^{\rho}_V \times \Theta$ and  found that the counterexample violation rate obtained is approximately  $1\%$ across all evaluation rounds with a tolerance of $10^{-3}$.  We next applied trajectory-based evaluation. For this, we simulated the closed-loop system under a parameter trajectory
\vspace{-2mm}
\begin{equation}\label{eq:quad-parameter-traj}
    \theta(t) =  \begin{bmatrix} 2 \sin(0.6 \pi t),  2 \cos(0.45\pi t),  2 \cos( 0.3 \pi t) \end{bmatrix}^T,
\end{equation}
which is depicted in \cref{fig:quad-dist-input} and simulates an external disturbance acting on the quadrotor in the inertial frame, 
and $10^6$ randomly sampled  initial states in  both $\Lambda^{\rho}_V$ and $\mathcal{X} \times \Theta$. \vspace{1mm} Although the parameter trajectory has a rate of variation much higher than the bounds of $\dot{\theta}(t)$ used in the training, the controller is able to handle it and yield satisfactory control performance. In all cases, the controller successfully drove the system to the equilibrium.  These results are summarized in \cref{tab:verification_results} and provide strong empirical support that 
$\Lambda^{\rho}_V$ serves as a subset of the R-ROA for the quadrotor system.

\begin{figure}[h]
\vspace{-3mm}
    \centering
\includegraphics[width=0.85\columnwidth]{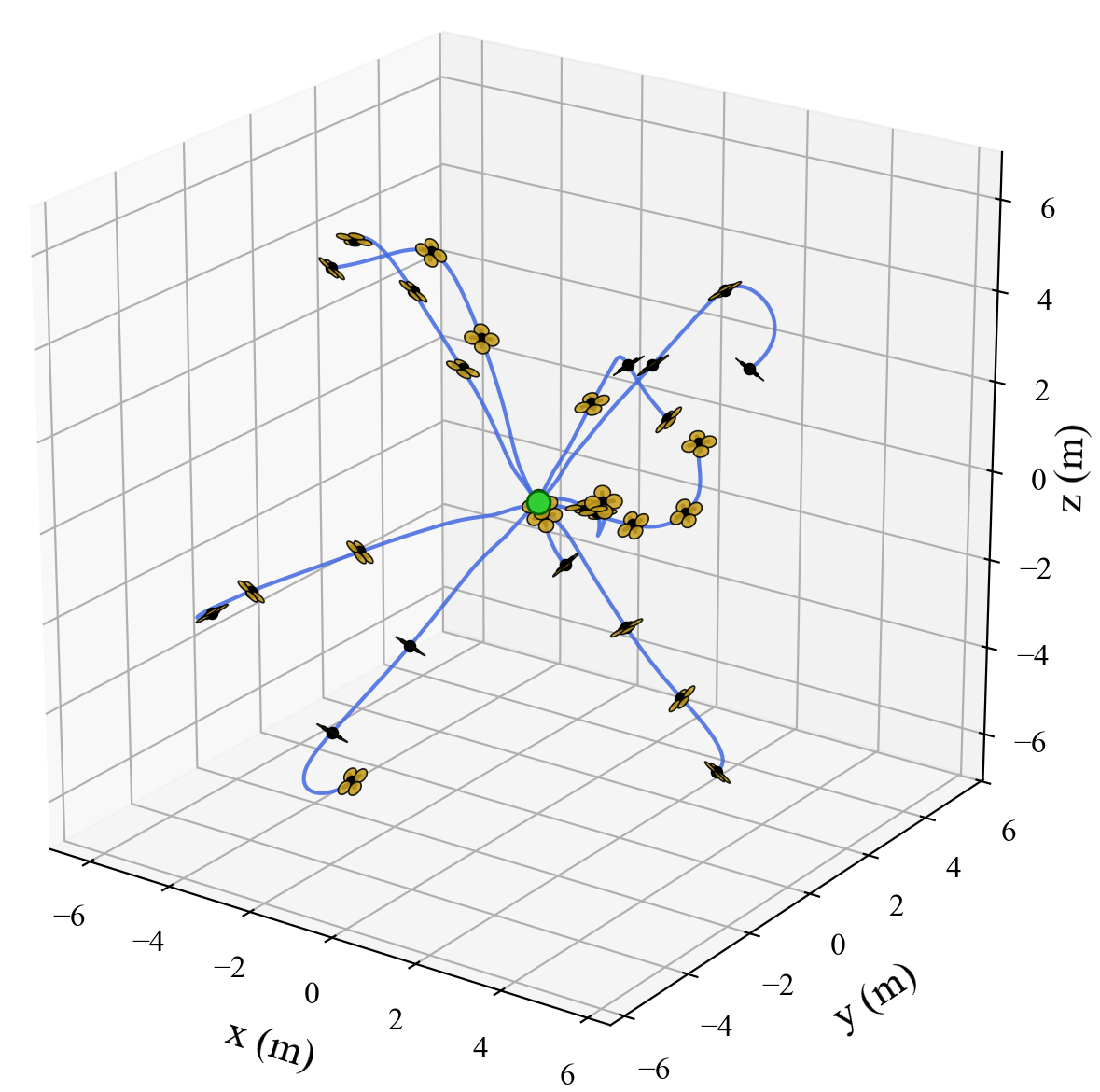} 
    \caption{  Closed-loop trajectories from 10 randomly sampled initial points under the parameter trajectory \cref{eq:quad-parameter-traj}.}
    
    \label{fig:quad-state}
\end{figure}

\begin{figure}[h]\
    \centering
\includegraphics[width=0.9\columnwidth]{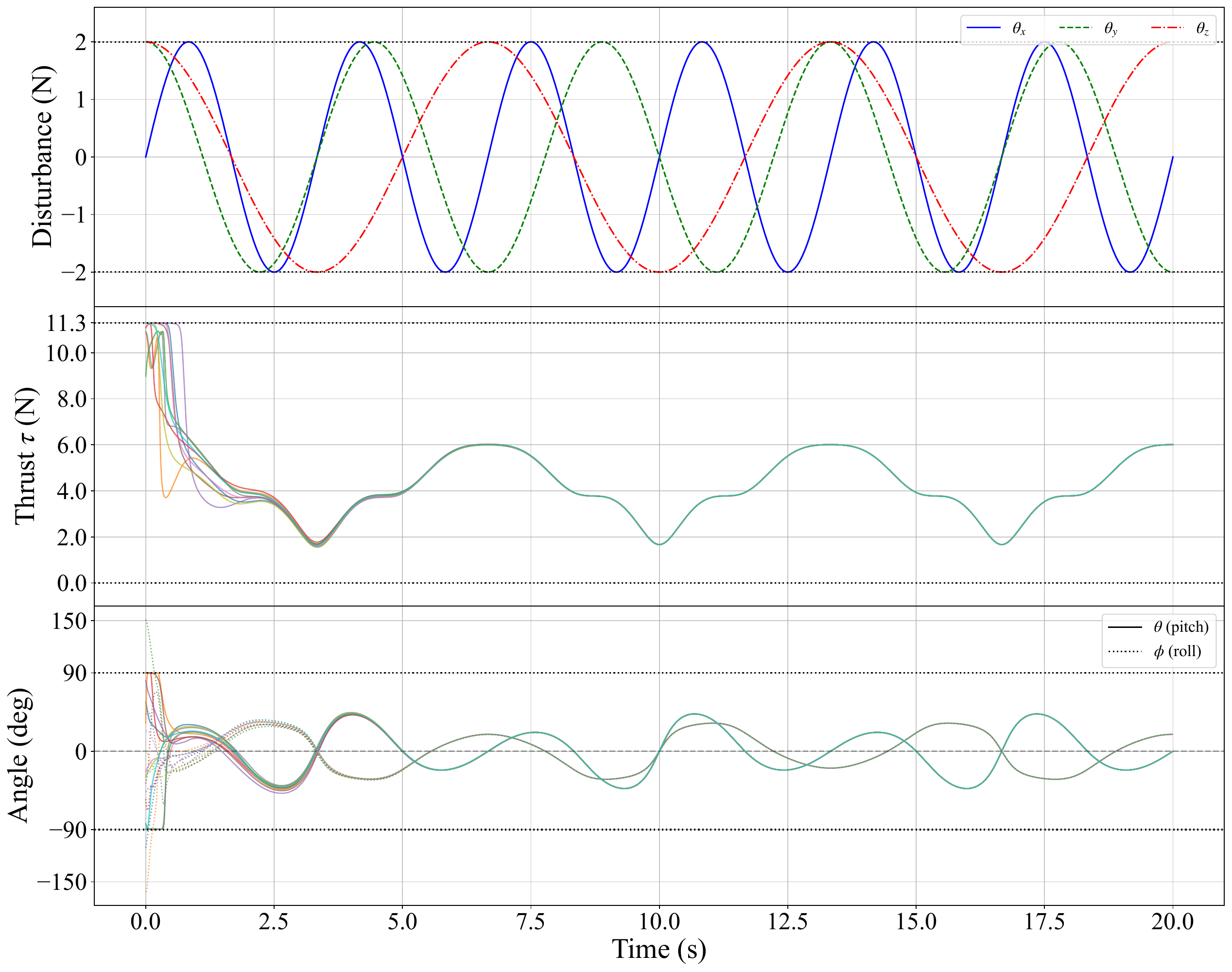} 
    \caption{ The parameter and control input trajectories of the quadrotor associated with \cref{fig:quad-state}. All the control inputs respect the input constraints (denoted by black dotted lines) explained in \cref{subsec:quad-dynam}.}
    \label{fig:quad-dist-input}
\end{figure}

\cref{fig:quad-state} shows the exemplary trajectories starting from {10} randomly sampled initial states, all of which converge to the equilibrium. As shown in \cref{fig:quad-dist-input}, after the states have already converged to the equilibrium in \cref{fig:quad-state}, the controller still constantly adjusts the control inputs to compensate for the external disturbances to keep the system at the  equilibrium.   

\section{Conclusions}\label{sec:conclusion}
We presented Neural-NPV, a learning-based framework to jointly synthesize  a PD Lyapunov function and PD controller for nonlinear parameter-varying systems. Our two-stage training framework uses a counterexample-guided training
procedure, and promotes the growth of the approximated robust region of attraction. The framework is applicable to control-non-affine systems, is scalable to high-dimensional systems, and handles input constraints seamlessly. 
Numerical experiments on an inverted pendulum and a quadrotor system demonstrate the effectiveness of our proposed approach. 

\bibliographystyle{ieeetr}
\bibliography{bib/refs-pan,bib/refs-new,bib/ref_niloy}

\end{document}

%% file: inc/packages.tex
\usepackage[fleqn]{amsmath}
\usepackage{amssymb,amsfonts,amsthm} 
\usepackage{tabularx}
\usepackage{mathabx}
\usepackage{graphicx,url} 
\usepackage{bm}
\usepackage{float}
\usepackage{makecell}
\let\labelindent\relax
\usepackage{enumitem}
\usepackage{subcaption}
\usepackage{wrapfig}
\usepackage{algorithm,algorithmicx} 
\usepackage{algpseudocode} 
\usepackage{multirow}
\usepackage{diagbox}
\renewcommand{\algorithmicrequire}{\textbf{Input:}}

\usepackage{cite}
\usepackage{soul}  
\usepackage[bookmarks=true]{hyperref}
\newcommand\rurl[1]{%
  \href{http://#1}{\nolinkurl{#1}}%
}

\usepackage{comment}
\usepackage{accents}
\newcommand{\ubar}[1]{\underaccent{\bar}{#1}}

\makeatletter
\def\cl@part {\@elt {chapter}}
\makeatother

\usepackage{cleveref}

\crefname{equation}{}{} 
\crefname{lemma}{Lemma}{Lemmas}
\crefname{theorem}{Theorem}{Theorems}
\crefname{table}{Table}{Tables}
\crefname{figure}{Fig.}{Figs.}
\crefname{remark}{Remark}{Remarks}
\crefname{assumption}{Assumption}{Assumptions}
\crefname{section}{Section}{Sections}
\crefname{definition}{Definition}{Definitions}
\crefname{algorithm}{Algorithm}{Algorithms}
\crefname{proposition}{Proposition}{Propositions}
\crefname{appendix}{Appendix}{Appendices}

%% file: inc/definition_this_paper.tex
\usepackage{accents}

\let\abs\relax
\newcommand{\abs}[1]{\left\lvert#1\right\rvert}
\newcommand{\norm}[1]{\left\lVert#1\right\rVert}

\def\mbR{\mathbb{R}}

\def\mcD{\mathcal{D}}

\def\mcX{\mathcal{X}}

\def\trieq{\triangleq}

\theoremstyle{definition}  \newtheorem{definition}{Definition}
\theoremstyle{definition} 
\newtheorem{assumption}{Assumption}
\theoremstyle{remark}  
\newtheorem{remark}{Remark}

\let\emptyset\varnothing

\makeatletter
\renewcommand*\env@matrix[1][\arraystretch]{%
  \edef\arraystretch{#1}%
  \hskip -\arraycolsep
  \let\@ifnextchar\new@ifnextchar
  \array{*\c@MaxMatrixCols c}}
\makeatother

\usepackage{commath}

\def \onetoN#1 {1,\dots,#1}

\def \nu {{n_u}}
\def \ntheta {{n_\theta}}

\def \cand {\textup{candidate}}
\def \nn {\textup{NN}}

%% file: bib/ref_niloy.bib
@string{ACC  = {American Control Conference}}

@article{dai2021lyapunov,
  title={Lyapunov-stable neural-network control},
  author={Dai, Hongkai and Landry, Benoit and Yang, Lujie and Pavone, Marco and Tedrake, Russ},
  journal={arXiv preprint arXiv:2109.14152},
  year={2021}
}

@article{li2026two,
  title={Two-Stage Learning of Stabilizing Neural Controllers via Zubov Sampling and Iterative Domain Expansion},
  author={Li, Haoyu and Zhong, Xiangru and Hu, Bin and Zhang, Huan},
  journal={NeurIPS},
  volume={38},
  pages={73980--74009},
  year={2026}
}

@inproceedings{zhou2022neural,
  title     = {Neural {Lyapunov} Control of Unknown Nonlinear Systems with Stability Guarantees},
  author    = {Zhou, Ruikun and Quartz, Thanin and De Sterck, Hans and Liu, Jun},
  booktitle = {NeurIPS},
  volume    = {35},
  pages     = {29113--29125},
  year      = {2022}
}

@inproceedings{liu2025formally,
  title     = {Formally Verified Physics-Informed Neural Control {Lyapunov} Functions},
  author    = {Liu, Jun and Fitzsimmons, Maxwell and Zhou, Ruikun and Meng, Yiming},
  booktitle = {Proc. ACC)},
  pages     = {1347--1354},
  year      = {2025},
  publisher = {IEEE}
}

@inproceedings{liu2024lyznet,
  title     = {{TOOL LyZNet}: A Lightweight {Python} Tool for Learning and Verifying Neural {Lyapunov} Functions and Regions of Attraction},
  author    = {Liu, Jun and Meng, Yiming and Fitzsimmons, Maxwell and Zhou, Ruikun},
  booktitle = {Proc. HSCC},
  articleno = {25},
  pages     = {1--8},
  year      = {2024},
  doi       = {10.1145/3641513.3650134}
}


%% file: bib/refs-pan.bib
@article{Rugh00gs_survey,
	title        = {Research on gain scheduling},
	author       = {Rugh, Wilson J and Shamma, Jeff S},
	year         = 2000,
	journal      = {Automatica},
	publisher    = {Elsevier},
	volume       = 36,
	number       = 10,
	pages        = {1401--1425}
}

@book{Moh12LPVBook,
	title        = {Control of {L}inear {P}arameter {V}arying {S}ystems with {A}pplications},
	author       = {Mohammadpour, Javad and Scherer, Carsten W},
	year         = 2012,
	publisher    = {Springer}
}

@article{tsukamoto2020neural-contraction,
	title        = {Neural contraction metrics for robust estimation and control: {A} convex optimization approach},
	author       = {Tsukamoto, Hiroyasu and Chung, Soon-Jo},
	year         = 2020,
	journal      = {IEEE Control Systems Letters},
	publisher    = {},
	volume       = 5,
	number       = 1,
	pages        = {211--216}
}

@article{chang2019neural-Lya,
	title        = {Neural {Lyapunov} Control},
	author       = {Chang, Ya-Chien and Roohi, Nima and Gao, Sicun},
	year         = 2019,
	journal      = {Advances in Neural Information Processing Systems},
	volume       = 32
}

@article{wu2025L1quad,
	title        = {{$\mathcal L_1$Quad: $\mathcal L_1$} Adaptive Augmentation of Geometric Control for Agile Quadrotors with Performance Guarantees},
	author       = {Z. Wu and S. Cheng and P. Zhao and others},
	year         = 2025,
	journal      = {IEEE TCST},
	volume       = 33,
	number       = 2,
	pages        = {597--612}
}

@inproceedings{prajna2004nonlinear-sos,
	title        = {{Nonlinear control synthesis by sum of squares optimization: A Lyapunov-based approach}},
	author       = {Prajna, Stephen and Papachristodoulou, Antonis and Wu, Fen},
	year         = 2004,
	booktitle    = {5th Asian control conference},
	volume       = 1,
	pages        = {157--165},
	organization = {IEEE}
}

@article{ahmadi2019dsos,
	title        = {{DSOS and SDSOS} optimization: more tractable alternatives to sum of squares and semidefinite optimization},
	author       = {Ahmadi, Amir Ali and Majumdar, Anirudha},
	year         = 2019,
	journal      = {SIAM Journal on Applied Algebra and Geometry},
	publisher    = {SIAM},
	volume       = 3,
	number       = 2,
	pages        = {193--230}
}

@article{dawson2023survey,
	title        = {Safe control With learned certificates: {A} survey of neural {L}yapunov, barrier, and contraction methods for robotics and control},
	author       = {Dawson, Charles and Gao, Sicun and Fan, Chuchu},
	year         = 2023,
	journal      = {IEEE T-RO},
	volume       = {39},
	number       = {3},
	pages        = {1749--1767},
	doi          = {10.1109/TRO.2022.3232542}
}

@book{khalil2002nonlinear-book,
	title        = {Nonlinear Systems},
	author       = {Hassan K. Khalil},
	year         = 2002,
	publisher    = {{Prentice Hall}},
	address      = {Englewood Cliffs, NJ}
}

@phdthesis{parrilo2000structured-sos,
	title        = {Structured Semidefinite Programs and Semialgebraic Geometry Methods in Robustness and Optimization},
	author       = {Parrilo, Pablo A},
	year         = 2000,
	school       = {Massachusetts Institute of Technology}
}

@article{zhao2023convex-cbf,
	title        = {Convex synthesis of control barrier functions under input constraints},
	author       = {Zhao, Pan and Ghabcheloo, Reza and Cheng, Yikun and Abdi, Hossein and Hovakimyan, Naira},
	year         = 2023,
	journal      = {IEEE Control Systems Letters},
	publisher    = {},
	volume       = 7,
	pages        = {3102--3107}
}

@article{yang2024lyapunov,
	title        = {Lyapunov-stable neural control for state and output feedback: {A} novel formulation for efficient synthesis and verification},
	author       = {Yang, Lujie and Dai, Hongkai and Shi, Zhouxing and Hsieh and others},
	year         = 2024,
	journal      = {arXiv preprint arXiv:2404.07956}
}

@article{zhao2025parameter-pd-clf,
	title        = {Parameter-Dependent Control {Lyapunov} Functions for Stabilizing Nonlinear Parameter-Varying Systems},
	author       = {Zhao, Pan},
	year         = 2025,
	journal      = {IEEE Control Systems Letters},
	volume       = 9,
	number       = {},
	pages        = {360--365},
	doi          = {10.1109/LCSYS.2025.3571820}
}

@article{lu2020domain-npv,
	title        = {On the domain of attraction and local stabilization of nonlinear parameter-varying systems},
	author       = {Lu, Linhong and Fu, Rong and Zeng, Jianping and Duan, Zhisheng},
	year         = 2020,
	journal      = {Int. J. Robust Nonlinear Control},
	publisher    = {},
	volume       = 30,
	number       = 1,
	pages        = {17--32}
}

@article{fu2018hinf-npv,
	title        = {{$H_\infty$} mixed stabilization of nonlinear parameter-varying systems},
	author       = {Fu, Rong and Zeng, Jianping and Duan, Zhisheng},
	year         = 2018,
	journal      = {International Journal of Robust and Nonlinear Control},
	publisher    = {Wiley Online Library},
	volume       = 28,
	number       = 17,
	pages        = {5232--5246}
}
